\documentclass[english]{article}
\usepackage[T1]{fontenc}
\usepackage[latin9]{inputenc}
\usepackage{geometry}
\usepackage{babel}
\usepackage{units}
\usepackage{tablefootnote}
\usepackage{amsmath}
\usepackage{amssymb}
\usepackage{cancel}
\usepackage{graphicx}
\usepackage[unicode=true]
 {hyperref}

\makeatletter

\providecommand{\tabularnewline}{\\}

\newcommand{\lyxaddress}[1]{
	\par {\raggedright #1
	\vspace{1.4em}
	\noindent\par}
}

\date{}
\usepackage{lineno, setspace,hyperref} 
\hypersetup{
colorlinks=true,
linkcolor=red,
citecolor = blue,
urlcolor=blue,
pdfpagemode=FullScreen}

\usepackage{xcolor} 

\@ifundefined{showcaptionsetup}{}{%
 \PassOptionsToPackage{caption=false}{subfig}}
\usepackage{subfig}
\makeatother

\begin{document}
\title{Finite Pulse Waves for Efficient Suppression of Evolving Mesoscale
Dendrites in Rechargeable Batteries}
\author{Asghar Aryanfar$^{\dagger\ddagger}$\thanks{Corresponding author. Email: \protect\href{http://aryanfar\%40caltech.edu}{aryanfar@caltech.edu}.},
Michael R. Hoffmann$^{\dagger}$ and William A. Goddard III$^{\dagger}$}
\maketitle

\lyxaddress{\begin{center}
\emph{$\dagger$ California Institute of Technology, 1200 E California
Blvd, Pasadena, CA 91125}\\
\emph{$\ddagger$ Bahçe\c{s}ehir University, 4 Ç\i ra\u{g}an Cad,
Be\c{s}ikta\c{s}, Istanbul, Turkey 34353}
\par\end{center}}
\begin{abstract}
The ramified and stochastic evolution of dendritic microstructures
has been a major issue on the safety and longevity of rechargeable
batteries, particularly for the utilization high-energy metallic electrodes.
We analytically develop criteria for the pulse characteristics leading
to the effective halting of the ramified electrodeposits grown during
extensive time scales beyond inter-ionic collisions. Our framework
is based on the competitive interplay between diffusion and electromigration
and tracks the gradient of ionic concentration throughout the entire
cycle of pulse-rest as a critical measure for heterogeneous evolution.
In particular, the framework incorporates the Brownian motion of the
ions and investigates the role of the geometry of the electrodeposition
interface. Our novel experimental observations verify the analytical
developments, where the the dimension-free developments allows the
application to the electrochemical systems of various scales. 
\end{abstract}

\section{Introduction}

Metallic anodes such as lithium, sodium and zinc are arguably highly
attractive candidates for use in high-energy and high-power density
rechargeable batteries.\cite{LI_14_REVIEW,Pei_14,SLATER_13} In particular,
lithium metal possess the lowest density and smallest ionic radius
which provides a very high gravimetric energy density and possesses
the highest electropositivity ($E^{0}=-3.04V$ vs SHE) that likely
provides the highest possible voltage, making it suitable for high-power
applications such as electric vehicles. ($\rho=0.53~g.cm^{-3}$).\cite{Li_19,XU_14}
During the charging, the fast-pace formation of microstructures with
relatively low surface energy from Brownian dynamics, leads to the
branched evolution with high surface to volume ratio.\cite{XU_04}
The quickening tree-like morphologies could occupy a large volume,
possibly reach the counter-electrode and short the cell. Additionally,
the can also dissolve from their thinner necks during subsequent discharge
period. Such a formation-dissolution cycle is particularly prominent
for the metal electrodes due to lack of intercalation\footnote{Intercalation: diffusion into inner layer as the housing for the charge,
as opposed to depositing in the surface.}.\cite{LI_14} Previous studies have investigated various factors
on dendritic formation such as current density,\cite{ORSINI_98} electrode
surface roughness \cite{MONROE_04,NIELSEN_15,Natisiavas_16}, impurities
\cite{HARRY_14,STEIGER_14}, solvent and electrolyte chemical composition
\cite{SCHWEIKERT_13,YOUNESI_15}, electrolyte concentration \cite{Brissot_99_2},
utilization of powder electrodes \cite{SEONG_08} and adhesive polymers\cite{STONE_12},
temperature \cite{ARYANFAR_15_2}, guiding scaffolds \cite{Yao_19,Qian_19},
capillary pressure \cite{Deng_19}, cathode morphology \cite{Abboud_19}
and mechanics \cite{XU_17,Wang_19}. Some of conventional characterization
techniques used include NMR \cite{BHATTACHARYYA_10} and MRI. \cite{CHANDRASHEKAR_12}
Recent studies also have shown the necessity of stability of solid
electrolyte interphase (i.e. SEI) layer for controlling the nucleation
and growth of the branched medium. \cite{Li_19_Energy,KASMAEE_16}

Earlier model of dendrites had focused on the electric field and space
charge as the main responsible mechanism \cite{CHAZALVIEL_90} while
the later models focused on ionic concentration causing the diffusion
limited aggregation (DLA). \cite{MONROE_03,Witten_83,Zhang_19} Both
mechanisms are part of the electrochemical potential \cite{Bard_80,Tewari_19},
indicating that each could be dominant depending on the localizations
of the electric potential or ionic concentration within the medium.
Nevertheless, their interplay has been explored rarely, especially
in continuum scale and realistic time intervals, matching scales of
the experimental time and space.

Dendrites instigation is rooted in the non-uniformity of electrode
surface morphology at the atomic scale combined with Brownian ionic
motion during electrodeposition. Any asperity in the surface provides
a sharp electric field that attracts the upcoming ions as a deposition
sink. Indeed the closeness of a convex surface to the counter electrode,
as the source of ionic release, is another contributing factor. In
fact, the same mechanism is responsible for the further semi-exponential
growth of dendrites in any scale. During each pulse period the ions
accumulate at the dendrites tips (unfavorable) due to high electric
field in convex geometry and during each subsequent rest period the
ions tend to diffuse away to other less concentrated regions (favorable).
The relaxation of ionic concentration during the idle period provides
a useful mechanism to achieve uniform deposition and growth during
the subsequent pulse interval. Such dynamics typically occurs within
the double layer (or stern layer \cite{BAZANT_11}) which is relatively
small and comparable to the Debye length. In high charge rates, the
ionic concentration is depleted and concentration on the depletion
reaches zero~\cite{Fleury_97}; Nonetheless, our continuum-level
study extends to larger scale, beyond the double layer region. \cite{ARYANFAR_15}

Pulse method has been qualitatively proved as a powerful approach
for the prevention of dendrites \cite{Li_01}, which has previously
been utilized for uniform electroplating.\cite{Chandrashekar_08}
In the preceding publication we have experimentally found that the
optimum rest period correlates well with the relaxation time of the
double layer for the blocking electrodes~\cite{ARYANFAR_14} which
is interpreted as the \emph{RC time} of the electrochemical system.
\cite{Bazant_04} We have explained qualitatively how relatively longer
pulse periods with identical duty cycles $D$ (or idle ratio $\gamma$)
will lead to longer and more quickening growing dendrites. We developed
coarse grained computationally affordable algorithm that allowed us
reach to the experimental time scale (\emph{$ms$}). Additionally,
in the recent theoretical work we indicated that there is an analytical
criterion for the optimal inhibition of growing dendrites. \cite{Aryanfar_18}

In this paper, we elaborate further in the range of acceptable duty
cycle $\mathbf{D}$ for suppression of stochastically-grown dendrites
and we develop new insight for the effective rest period on the curved
boundary. Subsequently we carry out experimental investigation to
verify our analytical developments on the pulse parameters. We perform
dimensional analysis to set our formulation applicable to the large
range of electrochemical devices.

\section{Methodology}

\begin{figure}
\noindent\begin{minipage}[t]{1\columnwidth}%
\begin{minipage}[t]{0.48\textwidth}%
\begin{center}
\includegraphics[height=0.24\textheight]{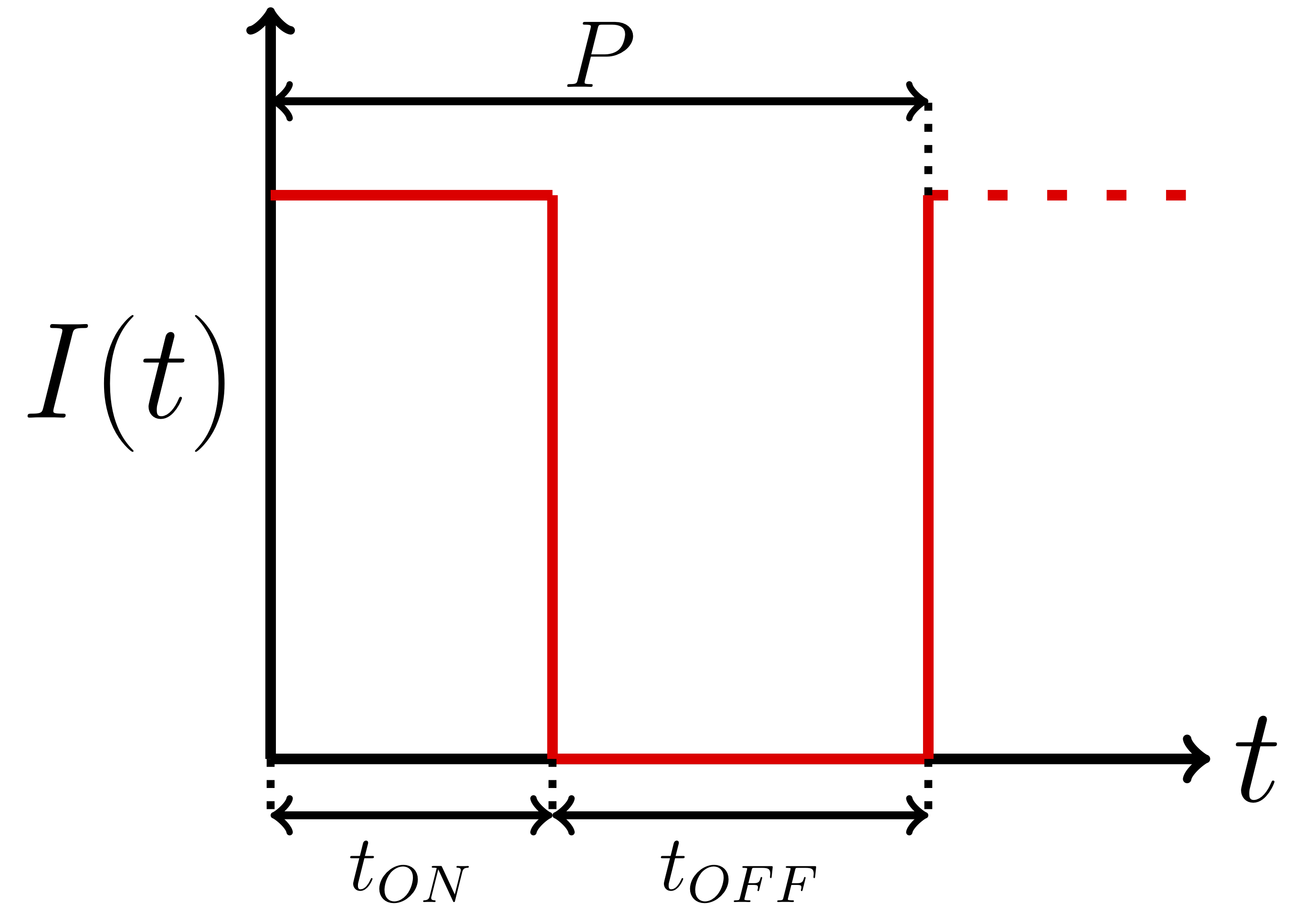}
\par\end{center}
\caption{Square pulse wave.\label{fig:PulseCurve}}
\end{minipage}\hfill{}%
\begin{minipage}[t]{0.48\textwidth}%
\begin{center}
\includegraphics[height=0.24\textheight]{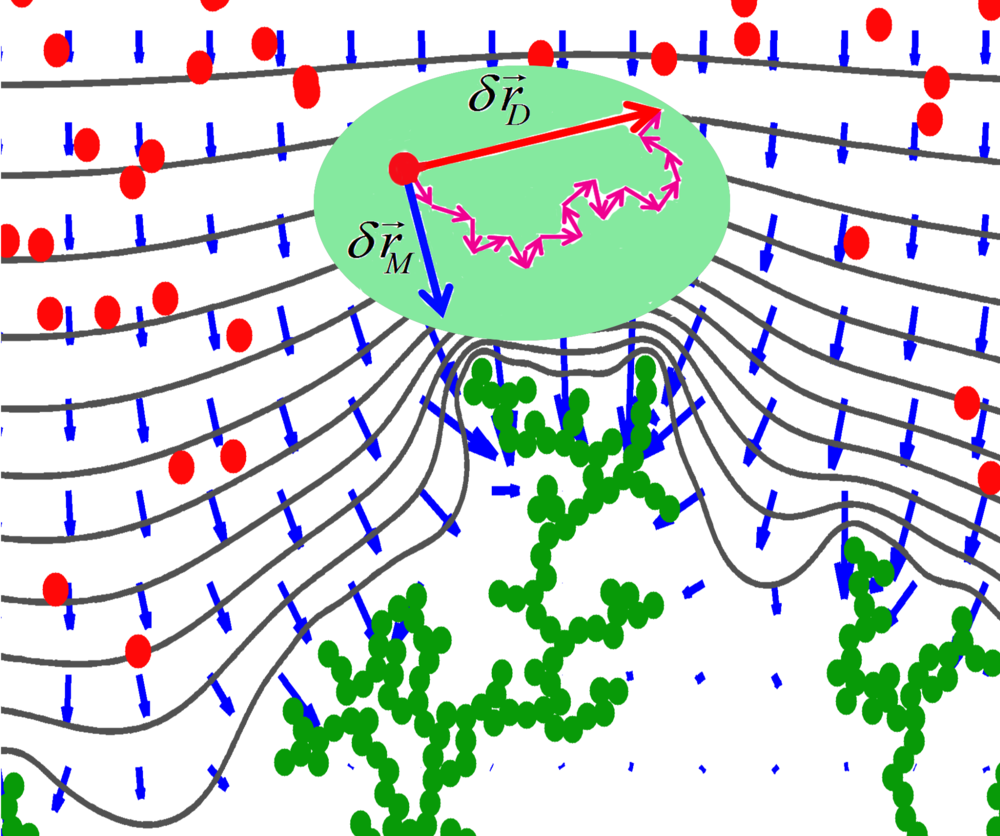}
\par\end{center}
\caption{The transport elements in the coarse scale of time and space.\label{fig:Displacements}}
\end{minipage}%
\end{minipage}
\end{figure}

The pulse charging in its simplest form consists of trains of square
active period $t_{ON}$ , followed by a square rest interval $t_{OFF}$
in terms of current $I$ or voltage $V$ as shown in Figure \ref{fig:PulseCurve}.
The period $P=t_{ON}+t_{OFF}$ is the time lapse of a full cycle.
Hence the pulse frequency $f$ is:

\begin{equation}
f=\frac{1}{t_{ON}+t_{OFF}}\label{eq:Frequency}
\end{equation}

and the duty cycle\textbf{ $\mathbf{D}$} represents the fraction
of time in the period $P$ that the pulse is active :

\begin{equation}
\mathbf{D}=ft_{ON}\label{eq:DftON}
\end{equation}

The electrochemical flux is generated either from the gradients of
concentration ($\nabla C$) or electric potential ($\nabla\phi$).
In the ionic scale, the regions of higher concentration tend to collide
and repel more and, given enough time, diffuse to lower concentration
zones, following Brownian motion. In the continuum scale, such inter-collisions
could be added-up and be represented by the diffusion length $\delta\vec{\textbf{r}}_{D}$
as: \cite{ARYANFAR_14}

\begin{equation}
\delta\vec{\textbf{r}}_{D}=\sqrt{2D^{+}\delta t}\text{ }\hat{\textbf{g}}\label{eq:DiffDis}
\end{equation}

where $\vec{\textbf{r}}_{D}$ is diffusion displacement of individual
ion, $D^{+}$ is the cationic diffusion coefficient in the electrolyte,
$\delta t$ is the coarse time interval\footnote{$\delta t=\sum_{i=1}^{n}\delta t_{i}$ where $\delta t_{k}$ is the
inter-collision time, typically in the range of $fs$.}, and $\hat{\textbf{g}}$ is a normalized vector in random direction,
representing the Brownian dynamics. The diffusion length represents
the average progress of a diffusive wave in a given time, obtained
directly from the diffusion equation. \cite{Philibert_06}

On the other hand, ions tend to acquire drift velocity in the electrolyte
medium when exposed to electric field and during the given time $\delta t$
their progress $\delta\vec{\textbf{r}}_{M}$ is given as:

\begin{equation}
\delta\vec{\textbf{r}}_{M}=\mu^{+}\vec{\textbf{E}}\delta t\label{eq:MigDis}
\end{equation}

where $\mu^{+}$ is the mobility of cations in electrolyte, $\vec{\textbf{E}}$
is the local electric field, which is the gradient of electric potential
($\vec{\textbf{E}}=-\nabla\phi$ ). Therefore the total effective
displacement $\delta\vec{\textbf{r}}$ with neglecting convection\footnote{Since Rayleigh number $Ra$ is highly dependent to the thickness (i.e.
$Ra\propto l^{3}$), for a thin layer of electrodeposition we have
$Ra<1500$ and thus the convection is negligible. \cite{Fox_16}} would be:

\begin{equation}
\delta\vec{\textbf{r}}=\delta\vec{\textbf{r}}_{D}+\delta\vec{\textbf{r}}_{M}\label{eq:TotalDis}
\end{equation}

as represented in the Figure \ref{fig:Displacements}. Based on the
Equations \ref{eq:Frequency} and \ref{eq:DftON}, defining two parameters
will uniquely characterize the pulse charge. We choose them as duty
cycle $\mathbf{D}$ (section \ref{subsec:OptimalD}) and the relaxation
(i.e. rest) period $t_{OFF}$ (section \ref{sec:OptRest}) as follows
next.

\subsection{Optimum Duty cycle \label{subsec:OptimalD}}

The vector sum in Equation \ref{eq:TotalDis} indicates that the diffusion,
as a random walk, can either contribute to electro-migration or prevents
its progress, depending on the local orientation of the gradients
of concentration and electric potential $\{\nabla C,\nabla\phi\}$.
From Figure \ref{fig:Displacements} it is visually obvious that the
sum of individual diffusional displacements after after the $n$ number
of collisions within the time interval $\delta t$ always is larger
(or equal to) than the on-step displacement of diffusion front during
the entire time coarse time interval ${\displaystyle \delta t=\sum_{i=1}^{n}\delta t_{i}}$
as:

\begin{equation}
\sum_{i=1}^{n}\sqrt{2D^{+}\delta t_{i}}\geq\sqrt{2D^{+}\sum_{i=1}^{n}\delta t_{i}}\label{eq:DIneq}
\end{equation}

We verify the Equation \ref{eq:DIneq} by induction. The equation
is obvious for value of $n:=1$, therefore we need to prove that if
Equation \ref{eq:DIneq} is true for the $n:=k$, then it should be
true for $n:=k+1$.

\begin{equation}
\sum_{i=1}^{k+1}\sqrt{2D^{+}\delta t_{i}}\geq\sqrt{2D^{+}\sum_{i=1}^{k+1}\delta t_{i}}\label{eq:k+1}
\end{equation}

Assuming that $\delta t_{i}=\delta t$ (i.e. equal segmentation) the
inequality \ref{eq:k+1} can be broken down as:

\[
\sum_{i=1}^{k}\sqrt{2D^{+}\delta t_{i}}+\sqrt{2D^{+}\delta t}\geq\sqrt{2D^{+}\sum_{i=1}^{k}\delta t_{i}+2D^{+}\delta t}
\]

Taking to the power 2, with simplification, we get the following:

\begin{equation}
2k(k+1)D^{+}\delta t\text{ }\geq0\text{ }\checkmark\label{eq:true}
\end{equation}

which means that Equation \ref{eq:k+1} is true for any consecutive
value of $k\rightarrow k+1$ and therefore indefinitely for any $k\in\mathbb{N}$.
In fact, Equation \ref{eq:DIneq} represents the extended version
of triangle inequality in terms of mean-square diffusion distance.
\cite{Khamsi_2011} During each pulse period $t_{ON}$, both diffusion
and migration are active for the ionic displacements. Therefore, depending
on their individual orientation they can help or hurt each other.
Thus the range of ionic displacement $|\delta\vec{\textbf{r}}|_{ON}$
in the pulse period is obtained as:

\begin{equation}
\mu^{+}\vec{\textbf{E}}\delta t-\sum_{i=1}^{n}\sqrt{2D^{+}\delta t_{i}}\leq|\delta\vec{\textbf{r}}|_{ON}\leq\mu^{+}\vec{\textbf{E}}\delta t+\sum_{i=1}^{n}\sqrt{2D^{+}\delta t_{i}}\label{eq:OnIneq}
\end{equation}

where $\mu^{+}$ and $D^{+}$ are the mobility and the diffusion coefficient
of local ions and $\vec{\textbf{E}}$ is the local electric field
respectively. For the Equation \ref{eq:OnIneq} to be valid, considering
Equation \ref{eq:DIneq}, one must have:

\begin{equation}
\mu^{+}\vec{\textbf{E}}\delta t-\sqrt{2D^{+}\delta t}\leq|\delta\vec{\textbf{r}}|_{ON}\leq\mu^{+}\vec{\textbf{E}}\delta t+\sqrt{2D^{+}\delta t}\label{eq:inequality}
\end{equation}

Such a random walk is succeeded with the idle period $t_{OFF}$ where
the the diffusion is the sole drive for the relaxation. In order to
have uniform electrodeposition, the average progress of diffusive
wave in the rest period $t_{OFF}$ has to be competitive enough with
the pulse interval $t_{ON}$, hence:

\begin{equation}
\sqrt{2D^{+}t_{OFF}}\geq\mu^{+}\vec{\textbf{E}}t_{ON}\pm\sqrt{2D^{+}t_{ON}}\label{eq:toffton}
\end{equation}

Without further look into Equation \ref{eq:toffton}, it is obvious
that $t_{OFF}\geq t_{ON}$. For simplification, we define the idle
ratio as $\gamma:={\displaystyle \frac{t_{OFF}}{t_{ON}}}$ and further
elaboration leads to:

\begin{equation}
\gamma\pm2\sqrt{\gamma}+1-\frac{\mu^{+}|\vec{\textbf{E}}|^{2}}{2RT}t_{ON}\geq0\label{eq:gamma}
\end{equation}

The solution to the Equation \ref{eq:gamma} represent the idle ratio
for effective fading of as:

\begin{equation}
\gamma\geq\left(1\pm|\vec{\textbf{E}}|\sqrt{\frac{2\mu^{+}t_{ON}}{RT}}\right)^{2}\label{eq:gammaProof}
\end{equation}
and the duty cycle $\mathbf{D}$ in term of the idle ratio $\gamma$
is obtained as:

\begin{equation}
\mathbf{D}=\frac{t_{ON}}{t_{ON}+t_{OFF}}=\frac{1}{1+\gamma}\label{eq:DutyGamma}
\end{equation}

Noting the Einstein relationship ($D^{+}=\mu^{+}RT$), the range of
acceptable duty cycle $\mathbf{D}$ would be: 
\begin{equation}
\mathbf{D}\leq\frac{1}{\left(1+{\displaystyle \frac{|\vec{\textbf{E}}|}{RT}}\sqrt{{\displaystyle \frac{D^{+}}{2f}}}\right)^{2}\pm1}\label{eq:Duty}
\end{equation}

\subsection{Optimum relaxation\label{sec:OptRest}}

The dendritic tip in fact attracts a significant number of ions due
to high electric field. Given sufficient time, such ionic concentration
profile can disappear in the vicinity of curved electrodeposits during
subsequent idle period. Therefore, the relaxation of concentration
plays a key role for preventing dendritic deposition. In fact the
oscillation of the concentration gradient repeatedly occurs during
each pulse-rest period.\cite{Fleury_97} Herein, we address a time
measure for concentration relaxation in the continuum scale with the
curved boundary rising from the tip of growing dendrites.

\begin{figure}
\centering{}\includegraphics[width=0.5\textwidth]{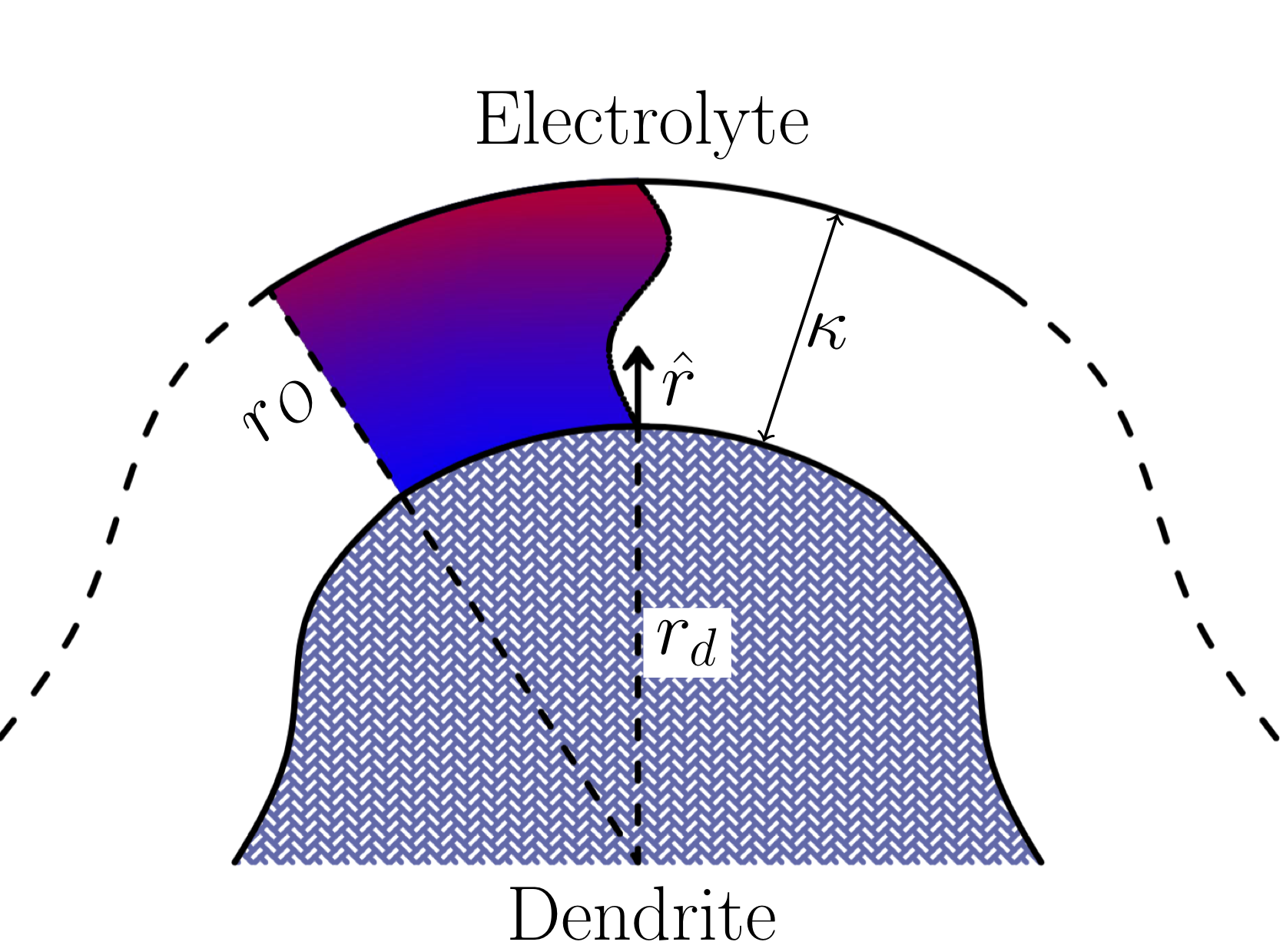}
\caption{The curved dendrites with the concentration gradient the vicinity
of surface.\label{fig:CurvedDendrite} }
\end{figure}

The schematics of the convex dendrites is shown in Figure \ref{fig:CurvedDendrite}
with the surrounding double layer of thickness of $\kappa$ and the
outer electro-neutral medium. The color gradient represents the concentration
profile in the double-layer region. The radius of curvature $r_{d}$
could vary from atomic radius ($r_{d}\approx r_{Li^{+}}\to10^{-9}m$)\cite{ARYANFAR_14}
to nearly flat surfaces ($r_{d}\rightarrow\infty$). Such a wide range
makes orders-of-magnitude of difference in the electric field and
concentration dynamics, making it critical factor to consider. We
define the normalized dendrite radial distance $\hat{r}\in[0,1]$
from the tip as: 
\begin{equation}
\hat{r}:=\frac{r-r_{d}}{\kappa}\label{eq:rhat}
\end{equation}

where $r$ is the center of curvature. Subsequently we can define
the normalized concentration $\hat{C}\in[0,1]$ as:

\begin{equation}
\hat{C}(\hat{r}):=\frac{C(\hat{r})}{C_{\infty}}\label{eq:chat}
\end{equation}

where the index $\infty$ represent the ambient electro-neutral medium.
The typical diffusion equation in polar 2D coordinates is defined
as: \footnote{The convection in the azimuthal direction $\hat{\theta}$ is neglected
due to below-threshold Rayleigh number.($Ra<1500$)}

\begin{align}
\frac{\partial C}{\partial t} & =\mathbf{\nabla}.(D^{+}\nabla C)\label{eq:RadDiffSup}\\
 & =\left(\frac{\partial}{\partial r}+\frac{1}{r}\frac{\partial}{\partial\theta}\right).\left(D(\frac{\partial C}{\partial r}+\frac{1}{r}\frac{\partial C}{\partial\theta})\right)\nonumber \\
 & =D^{+}\left(\frac{\partial^{2}C}{\partial r^{2}}+\frac{1}{r}\frac{\partial C}{\partial r}\right)\nonumber 
\end{align}

Using the chain derivative and noting Equation \ref{eq:chat} we get:

\begin{equation}
\frac{\partial C}{\partial t}=\frac{\partial C}{\partial\hat{C}}\frac{\partial\hat{C}}{\partial t}=C_{\infty}\frac{\partial\hat{C}}{\partial t}\label{eq:CChat_t}
\end{equation}

Respectively for the radial space derivative is obtained, considering
Equation \ref{eq:rhat} as:

\begin{equation}
\frac{\partial C}{\partial r}=\frac{\partial C}{\partial\hat{C}}\frac{\partial\hat{C}}{\partial\hat{r}}\frac{\partial\hat{r}}{\partial r}=\frac{C_{\infty}}{\kappa}\frac{\partial\hat{C}}{\partial\hat{r}}\label{eq:CChat_r}
\end{equation}

The second radial derivative is respectively obtained as:

\begin{align}
\frac{\partial^{2}C}{\partial r^{2}} & =\frac{\partial}{\partial r}\left(\frac{\partial C}{\partial r}\right)=\frac{\partial}{\partial r}\left(\frac{C_{\infty}}{\kappa}\frac{\partial\hat{C}}{\partial\hat{r}}\right)\label{eq:CChat_r2}\\
 & =\frac{C_{\infty}}{\kappa}\frac{\partial}{\partial\hat{r}}\left(\frac{\partial\hat{C}}{\partial\hat{r}}\right)\frac{\partial\hat{r}}{\partial r}=\frac{C_{\infty}}{\kappa^{2}}\frac{\partial^{2}\hat{C}}{\partial\hat{r}^{2}}\nonumber 
\end{align}

Therefore, replacing all the obtained terms from Equations \ref{eq:CChat_t},
\ref{eq:CChat_r} and \ref{eq:CChat_r2} into Equations \ref{eq:RadDiffSup}
and simplification leads to:
\begin{center}
\begin{equation}
\kappa^{2}\frac{\partial\hat{C}}{\partial t}=D^{+}\left(\frac{\partial^{2}\hat{C}}{\partial\hat{r}^{2}}+\frac{\kappa}{r_{d}+\kappa\hat{r}}\frac{\partial\hat{C}}{\partial\hat{r}}\right)\label{eq:Laplacian}
\end{equation}
\par\end{center}

Regarding the boundary conditions, while the concentration is depleted
in the double layer, in the outer boundary ($\hat{r}\rightarrow1$)
it remains as the ambient value $C_{\infty}$ :

\begin{equation}
\hat{C}(1,t)=1\label{eq:BCOuter}
\end{equation}

During the charge period, a constant reduction ionic flux $j$ is
fed to the dendrite and respectively during the idle period there
will be no reaction since the dendrites will not accept any ions.
Therefore:

\begin{equation}
\left\{ \begin{array}{ll}
\cfrac{\partial\hat{C}}{\partial\hat{r}}(0,t) & ={\displaystyle \frac{\kappa}{C_{\infty}}\frac{\partial C}{\partial r}=-\frac{\kappa j}{C_{\infty}D^{+}}\text{\ensuremath{\hspace{0.9cm}\text{Pulse}}}}\\
\cfrac{\partial\hat{C}}{\partial\hat{r}}(0,t) & =0\hspace{3.85cm}\text{Rest}
\end{array}\right.\label{eq:BCs}
\end{equation}
The Equation \ref{eq:Laplacian} can be solved numerically using a
finite difference method where the $\hat{C}_{i}^{j}$ represents the
concentration in the radial direction $\hat{r}(i)$ and at the time
$t(j)$. Performing segmentation in the time $\delta t$ and space
$\delta\hat{r}$ and utilizing the scheme of forward-move in time
and space ($FTFS$), we arrive at the following:

\begin{equation}
\frac{\hat{C}_{i}^{j+1}-\hat{C}_{i}^{j}}{\delta t}=D^{+}\left(\frac{1}{\kappa^{2}}\frac{\hat{C}_{i+1}^{j}-2\hat{C}_{i}^{j}+\hat{C}_{i-1}^{j}}{\delta\hat{r}^{2}}+\frac{1}{\kappa\left(r_{d}+\kappa\hat{r}\right)}\left(\frac{\hat{C}_{i+1}^{j}-\hat{C}_{i}^{j}}{\delta\hat{r}}\right)\right)\label{eq:DiffDiscrete}
\end{equation}
The Equation \ref{eq:DiffDiscrete} can be rearranged in terms of
individual concentration terms as:

\begin{equation}
C_{i}^{j+1}=\left(1-\frac{2}{\kappa^{2}}\cfrac{D^{+}\delta t}{\delta\hat{r}^{2}}-\frac{D^{+}\delta t}{\left(r_{d}+\kappa\hat{r}\right)\kappa\delta\hat{r}}\right)C_{i}^{j}+\left(\frac{D^{+}\delta t}{\kappa^{2}\delta\hat{r}^{2}}+\frac{D^{+}\delta t}{(r_{d}+\kappa\hat{r})\kappa\delta\hat{r}}\right)\hat{C}_{i+1}^{j}+\frac{1}{\kappa^{2}}\cfrac{D^{+}\delta t}{\delta\hat{r}^{2}}\hat{C}_{i-1}^{j}\label{eq:OpenDiscrete}
\end{equation}

which can be simplified to as the following:

\begin{equation}
\hat{C}_{i}^{j+1}=\left(1-\frac{2Q_{1}}{\delta\hat{r}^{2}}-\frac{\hat{r}}{\delta\hat{r}}Q_{2}\right)\hat{C}_{i}^{j}+\left(Q_{1}+\frac{\hat{r}}{\delta\hat{r}}Q_{2}\right)\hat{C}_{i+1}^{j}+Q_{1}\hat{C}_{i-1}^{j}\label{eq:FinDifArr}
\end{equation}

The terms $Q_{1}$ and $Q_{2}$ are the dimension-free quotients,
as below:

\begin{equation}
\left\{ \begin{array}{ll}
{\displaystyle Q_{1}=\frac{D^{+}\delta t}{\kappa^{2}}}\\
{\displaystyle {\displaystyle Q_{2}=\frac{D^{+}\delta t}{(r_{d}+\kappa\hat{r})\kappa\hat{r}}}}
\end{array}\right.\label{eq:Qoutients}
\end{equation}

Equation~\ref{eq:FinDifArr} should possess enough resolution in
time $\delta t$ to capture the variations in space $\delta\hat{r}$.
Therefore the stability criterion requires for the coefficient of
$\hat{C}_{i}^{j}$ to be non-negative:

\begin{equation}
\delta\hat{r}^{2}-\frac{D^{+}\delta t}{(r_{d}+\kappa\hat{r})\kappa}\delta\hat{r}-{\displaystyle \frac{2D^{+}\delta t}{\kappa^{2}}}\geq0\label{eq:Parabolic}
\end{equation}

this is a parabolic equation in terms of $\delta\hat{r}$. Therefore
noting $\hat{r}_{max}=1$, we have:

\begin{equation}
\delta\hat{r}\geq{\displaystyle \frac{D^{+}\delta t}{2\kappa(r_{d}+\kappa)}}\pm\frac{1}{2}\sqrt{\left({\displaystyle \frac{D^{+}\delta t}{\kappa(r_{d}+\kappa)}}\right)^{2}+8\frac{D^{+}\delta t}{\kappa(r_{d}+\kappa)}}\label{eq:drdtIneq}
\end{equation}

Looking closer to the term ${\displaystyle 8\frac{D^{+}\delta t}{\kappa(r_{d}+\kappa)}}$,
the nominator in fact represents the square of the progress for the
diffusive wave during the time $\delta t$, which, in order to be
captured, must fall inside the double layer, with the scale of $\kappa$.
In other words: $D^{+}\delta t\leq\kappa^{2}<\kappa(r_{d}+\kappa)\Rightarrow{\displaystyle \frac{D^{+}\delta t}{\kappa(r_{d}+\kappa)}<1}$
. Thus in order for the Equation \ref{eq:drdtIneq} to be true, simplification
of RHS will lead to:  

\begin{equation}
\delta\hat{r}\geq\frac{2D^{+}\delta t}{(r_{d}+\kappa)\kappa}\label{eq:SpaceTime}
\end{equation}

where $r_{d}+\kappa:=r_{O}$ from Figure \ref{fig:CurvedDendrite}.
The Equation \ref{eq:SpaceTime} in fact inherits the scale-free time
measure for concentration dynamics as:

\begin{equation}
{\displaystyle \delta\hat{t}}={\displaystyle \frac{D^{+}\delta t}{\kappa(\kappa+r_{d})}}\label{eq:TimeMeasure}
\end{equation}

and the dimension-free space-time criterion for all $\hat{r}\in[0,1]$
is obtained as:

\begin{equation}
\frac{\delta\hat{t}}{\delta\hat{r}}<\frac{1}{2}\label{eq:DFreeTFree}
\end{equation}

\begin{figure}
\centering{}\includegraphics[width=0.5\textwidth]{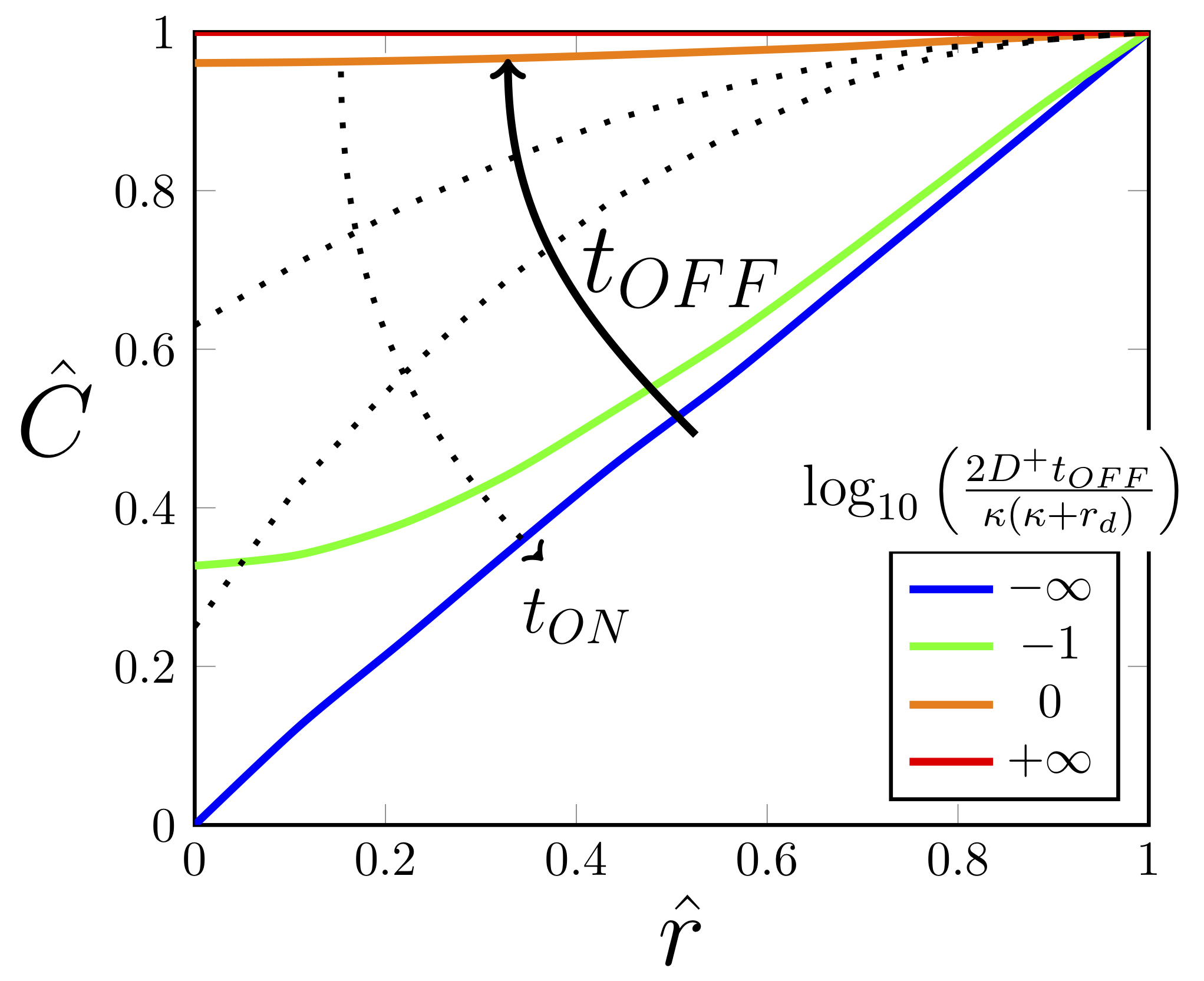}\caption{The concentration profile during the rest period $t_{OFF}$ (color)
and the subsequent pulse interval $t_{ON}$ (dots). \label{fig:CProfile}}
\end{figure}

The evolution of concentration profile $\hat{C}$ from Equation \ref{eq:Laplacian}
during the entire cycle of pulse-rest has been shown in Figure \ref{fig:CProfile}
with the constants given in the Table \ref{tab:SimPars}.

\begin{table}
\begin{centering}
\begin{tabular}{ccccc}
\hline 
\multicolumn{1}{|c|}{Parameter} & \multicolumn{1}{c|}{$D^{+}$} & \multicolumn{1}{c|}{$\kappa$} & \multicolumn{1}{c|}{$j$} & \multicolumn{1}{c|}{$r_{d}$}\tabularnewline
\hline 
Value & $2.58\times10^{-10}$ & $20$ & $10^{-4}$ & $20$\tabularnewline
Unit & $m^{2}/s$ & $\mu m$ & $C/m^{2}s$ & $nm$\tabularnewline
Ref. & \cite{Brissot_99_2} & \cite{CHAZALVIEL_90} & \cite{Barton_62} & \cite{Barton_62}\tabularnewline
\end{tabular}
\par\end{centering}
\caption{Simulation Parameters. \label{tab:SimPars}}
\end{table}

\section{Experimental}

\begin{figure}
\subfloat[Naked-eye observation of dendrites.\label{fig:Naked}\cite{ARYANFAR_15_2}]{\noindent \begin{centering}
\includegraphics[height=0.18\textheight]{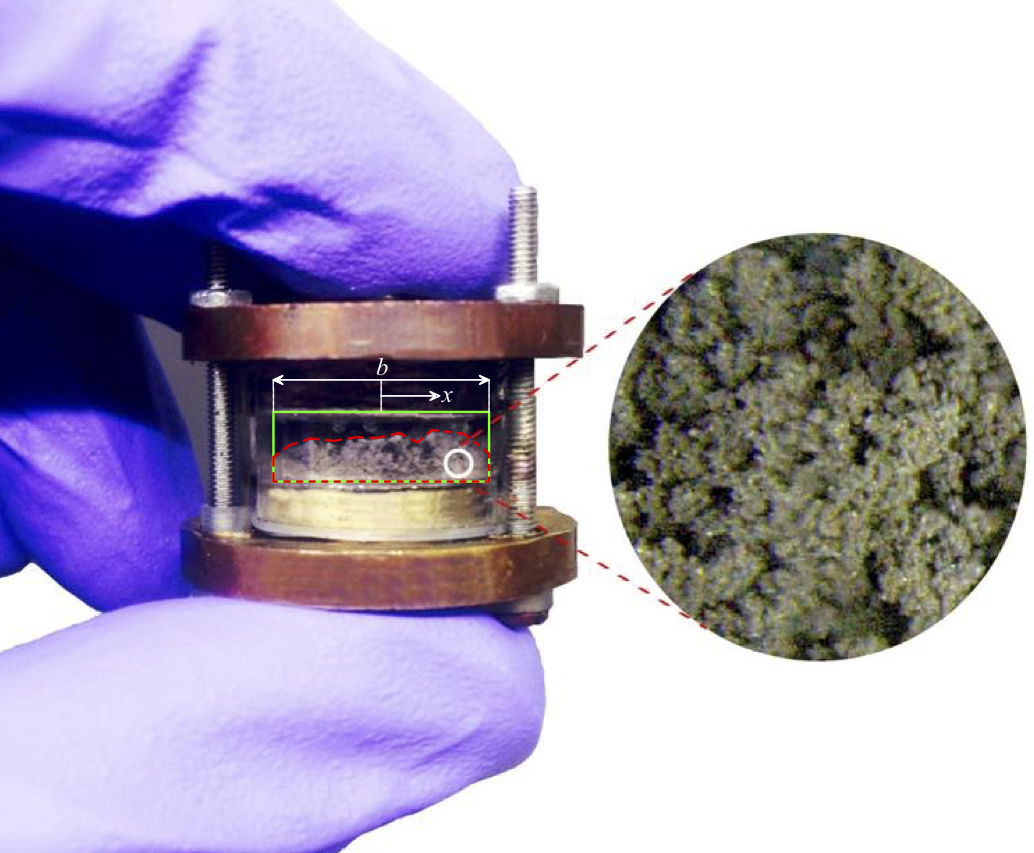}
\par\end{centering}
}\hfill{}\subfloat[Sample tracking of the suppression for $f=100Hz$. \label{fig:ExpSamples}]{\noindent \begin{centering}
\includegraphics[height=0.18\textheight]{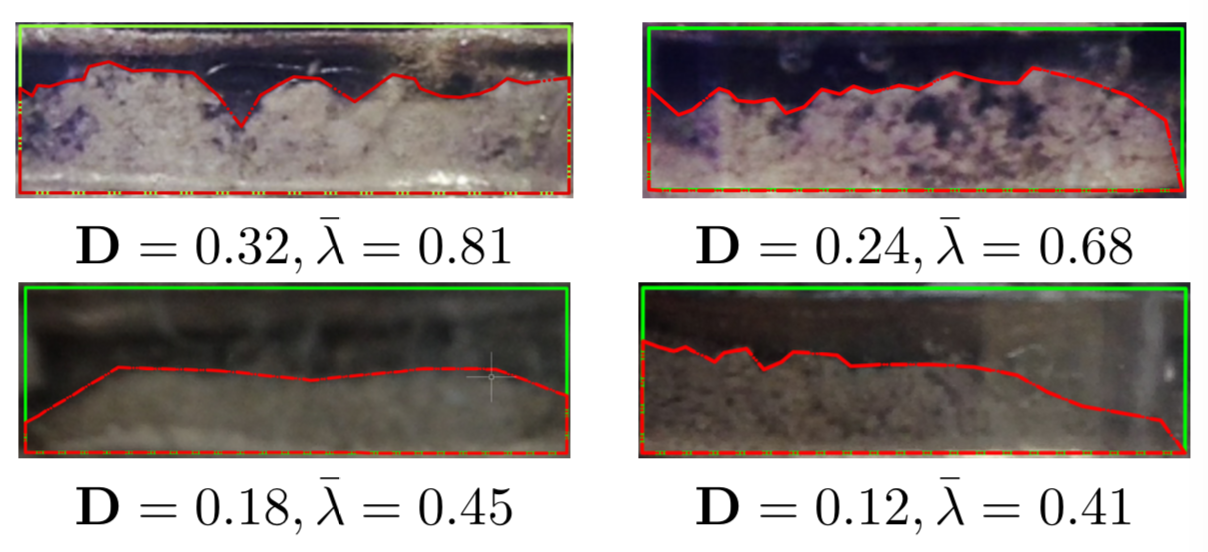}
\par\end{centering}
}

\hfill{}\subfloat[Extracting the dendrite measure from experimental images. \label{fig:ExpProcedure}]{\begin{centering}
\includegraphics[width=1\textwidth]{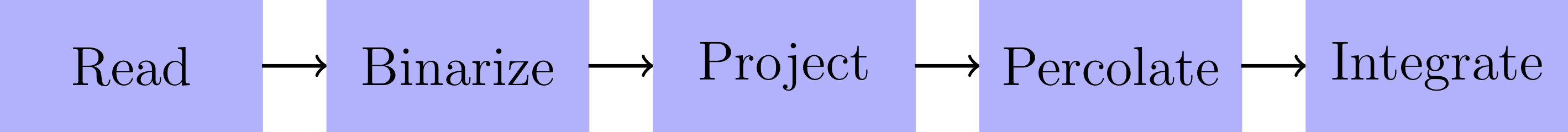}
\par\end{centering}
}

\caption{Experimental Procedure.}
\end{figure}

The dendritic measurements has been carried out in a manually-fabricated
electrolytic cell\cite{ARYANFAR_17}, that provides the possibility
of \emph{in-situ} observation of growing dendrites from the periphery
in real-time as shown in Figure \ref{fig:Naked}. The sandwich cell
consists of two $Li^{0}$ foil disc electrodes ($D=1.59cm$) with
the inter-electrode separation of $L=0.32cm$ by means of a transparent
acrylic PMMA housing. The fabricated cells were filled with $0.4cm^{3}$
of LiPF$_{6}$ in a the stoichiometric compound of EC:EMC$\equiv$1:1.
We performed the operations in an argon-filled glovebox($H_{2}O,O_{2}<0.5ppm$)
. Multiple such cells were electrolyzed with current density pulse
trains consisting of a range of frequencies $f$, generated by a programmable
multichannel charger. After the passage of $48mAh$ ($\approx173C$)
through the cells, 3 images within the periphery of $120^{0}$ were
taken by means of Leica M205FA optical microscope through the acrylic
separator. The image processing algorithm given in the Figure \ref{fig:ExpProcedure}
is described as below:

1. The RGB image is read to the program by 3 values of $\{R,G,B\}\in[0,255]$
and has been converted to a grayscale image $I$ with individual values
of range $I_{i,j}\in[0,1]$.

2. The image is binarized from Otsu's method. For this purpose a critical
grayness threshold $I_{c}$ has been chosen to approximate the grayscale
image $I$ with a binarized image $J$ as below:

\[
J_{i,j}=\begin{cases}
1 & I_{i,j}\geq I_{c}\\
0 & I_{i,j}<I_{c}
\end{cases}
\]

the threshold value $I_{c}$ has been chosen to minimize the weighted
intra-class variance $\sigma^{2}$ defined as:

\[
\begin{cases}
\sigma^{2}=\omega_{0}\sigma_{0}^{2}+\omega_{1}\sigma_{1}^{2}\\
\omega_{0}+\omega_{1}=1
\end{cases}
\]

where $\omega_{0}$ and $\omega_{1}$ are the total fraction of element
divided by the value of $I_{c}$ and $\sigma_{0}^{2}$ and $\sigma_{1}^{2}$
are their respective variances.\cite{Otsu_75}.

3. The circular sandwich cell with the radius $R$ has been divided
of 3 arcs with the angle of ${\displaystyle \frac{2\pi}{3}}$ and
width incremental length of $\delta x$, which is supposed to be projected
to a 2D plane with the incremental width of $\delta x'$. From Figure
\ref{fig:Naked} due to geometry we have: $x={\displaystyle \frac{D}{2}}sin(\theta)$,
$\rightarrow$ $dx={\displaystyle \frac{D}{2}}cos(\theta)d\theta$,
where $cos(\theta)=\sqrt{{\displaystyle 1-\frac{4x^{2}}{D^{2}}}}$
; hence:

\[
\delta x'=\frac{\delta x}{\sqrt{1-{\displaystyle \frac{4x^{2}}{D^{2}}}}}
\]

where $D$ is diameter of the cell. \cite{ARYANFAR_14_2}

4. Starting from the electrode surface, the occupied space by the
dendrites has been calculated by the square site percolation paradigm.
\cite{Aryanfar_19}

5. The infinitesimal calculations have been integrated and normalized
to inter-electrode distance (${\displaystyle \hat{\lambda}_{i}:=\lambda_{i}/l}$
) to get the dendrite measure $\bar{\lambda}$, as shown in Figure
\ref{fig:ExpProcedure} as:
\begin{align}
\bar{\lambda} & =\frac{1}{\pi Dl}\sum_{k=1}^{3}\int_{-\frac{\pi}{3}}^{+\frac{\pi}{3}}\hat{\lambda}_{k}(\theta){\displaystyle \frac{D}{2}}d\theta\label{eq:Lambda}\\
 & =\frac{1}{\pi Dl}\sum_{k=1}^{3}\int_{-\frac{\pi}{3}}^{+\frac{\pi}{3}}\frac{\hat{\lambda}_{k}(x)dx}{\sqrt{1-{\displaystyle \frac{4x^{2}}{D^{2}}}}}\nonumber 
\end{align}
The integral Equation \ref{eq:Lambda} has been obtained by incremental
sum from experimental data. The optimal duty cycle $\mathbf{D}$ has
been considered where the sensitivity of dendrites metric $\bar{\lambda}$
to duty cycle $\mathbf{D}$ is less than $10\%$. Hence:

\[
\mathbf{D_{opt}}:\equiv\left\{ \frac{\Delta\bar{\lambda}}{\Delta\mathbf{D}}\leq0.1\right\} 
\]

Figure \ref{fig:ExpSamples} shows such investigation for the sample
pulse frequency of $f=100Hz$. \footnote{While the resolution of some images would not be quite high due to
observation conditions from post-experiment acrylic separator, they
suffice for binarization purpose shown in Figure \ref{fig:ExpProcedure}
and extracting the figure of merit $\bar{\lambda}$.}The experimental parameters for further data are given in the Table
\ref{tab:ExpParameters}. \footnote{Note that the current density $i$ and the ionic flux $j$ are correlated
with $i=zFj$, where $z$ is the valence number of charge carriers
and $F=96.5\text{ }\nicefrac{kC}{mol}$ is the Faraday's constant,
representing the amount of charge per mole.}

\begin{table}
\begin{centering}
\begin{tabular}{cccccccc}
\hline 
\multicolumn{1}{|c|}{Parameter} & \multicolumn{1}{c|}{$f$} & \multicolumn{1}{c|}{$i$} & \multicolumn{1}{c|}{$l$} & \multicolumn{1}{c|}{$R$} & \multicolumn{1}{c|}{$T$} & \multicolumn{1}{c|}{$C_{\infty}$} & \multicolumn{1}{c|}{$|\mathbf{\vec{E}}|_{max}$\tablefootnote{The maximum value of electric potential ($|\mathbf{\vec{E}}|_{max}$)
is far more than the average electric filed in the inter-electrode
space, due to the closer proximity of the dendritic microstructures
to the counter-electrode, as well as the extremely high curvature
of the dendrite, reachable to atomic scale. i.e. $|\mathbf{\vec{E}}|_{max}\gg{\displaystyle \frac{\Delta V}{l}}$}}\tabularnewline
\hline 
Value & $\{25,40,100,250,1000\}$ & $1$ & $3.175$ & $0.795$ & $298$ & $1$ & $10^{8}$\tabularnewline
Unit & $mHz$ & $mA/cm^{2}$ & $mm$ & $cm$ & $K$ & $M$ & $N/m$\tabularnewline
\end{tabular}
\par\end{centering}
\caption{Experimental Parameters. \label{tab:ExpParameters}}
\end{table}

\section{Results and discussion}

\subsection{Duty Cycle}

\begin{figure}
\centering{}\includegraphics[height=0.5\textwidth]{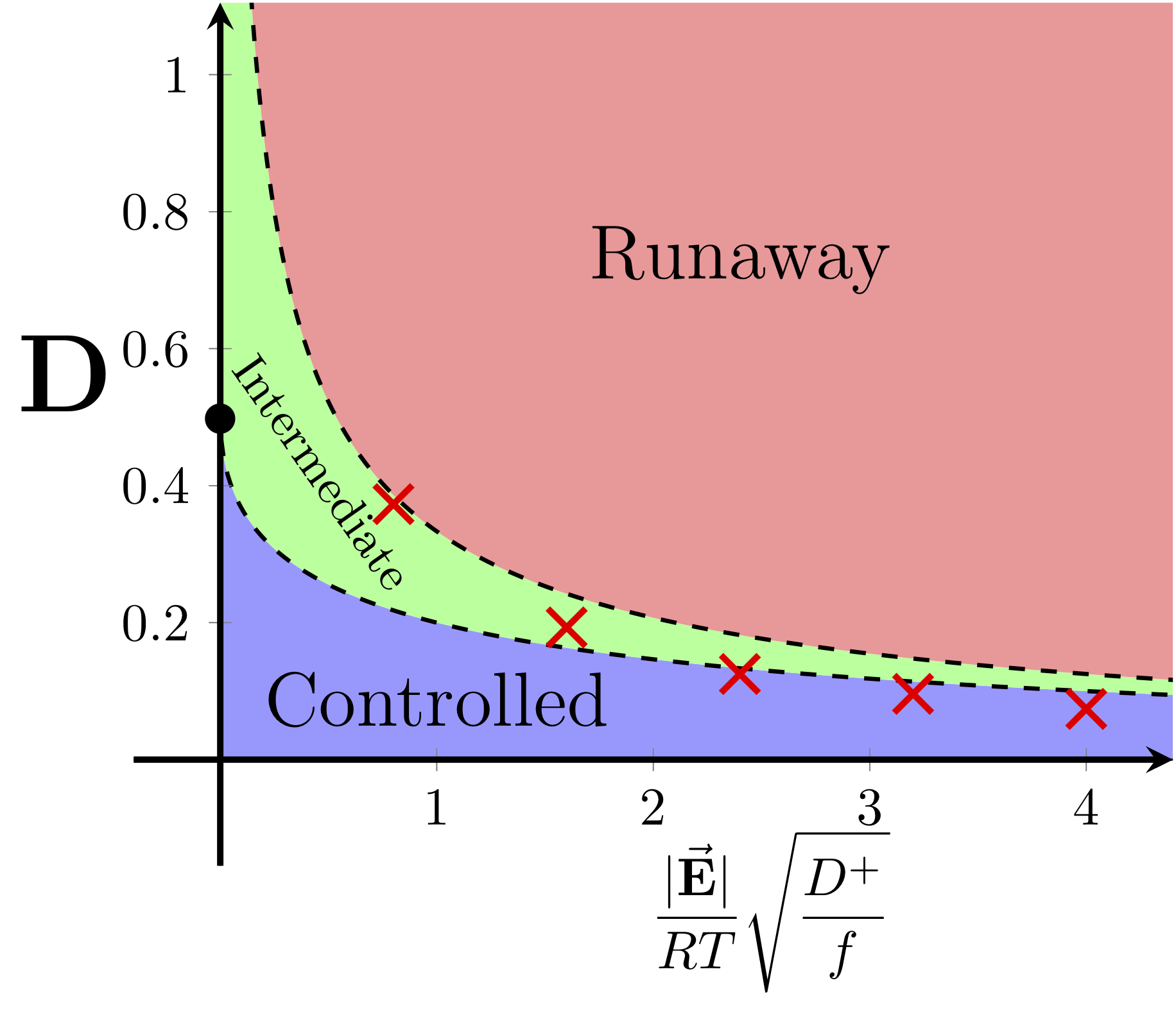}\caption{The regimes based on duty cycle $\mathbf{D}$ and frequency $f$ showing
the safe/unsafe charging zones. $\times$: Experimental data , $\bullet$:
Theoretical limit. \label{fig:SafeUnsafe}}
\end{figure}

Figure \ref{fig:SafeUnsafe} visualizes the range of acceptable duty
cycle $\mathbf{D}$ for the suppression of microstructures. In fact
its theoretical limit can be obtained when the pulse frequency $f$
is increased indefinitely as: \footnote{Note that the lower bound has been considered for the inequality to
be true in all instances.}

\[
\mathbf{D}_{max}=\lim_{f\to\infty}\frac{1}{\left(1+{\displaystyle \frac{|\vec{\textbf{E}}|}{RT}}\sqrt{{\displaystyle \frac{D^{+}}{2f}}}\right)^{2}+1}=\frac{1}{2}
\]

Additionally in the Figure \ref{fig:SafeUnsafe}, the \emph{Controlled}
region shows the safe zone for pulse charging where the ionic progress
in the idle period is competitive enough with the pulse duration.
Vice versa, the \emph{Runaway }region represents the regime where
the average ionic lead during pulse wave takes over the rest period,
and therefore the dendritic growth would be exacerbated. Nonetheless,
the \emph{Intermediate} region shows the role of random walk where
the certainty is less than the other areas. The experimental observations
in this Figure also illustrate a very high agreement with the analytical
trend.

Additionally, it is obvious that the pulse duty cycle $\mathbf{D}$
correlates inversely with the diffusion coefficient $D^{+}$ and to
a higher extend to the magnitude of the electric field $|\vec{\textbf{E}}|$.
Both parameters exacerbate the growth kinetics and in trade-off, the
duty cycle would have to become more conservative. In fact, the augmentation
of electric field in the dendritic tips during the real-time growth
causes the quickening growth behavior, which has been addressed before.
\cite{Aryanfar_18,MONROE_03}

\subsection{Concentration profile}

Looking closer to the depletion-accrual cycle of concentration during
full pulse-rest period shown in Figure \ref{fig:CProfile}, we have
the following inequality:

\begin{equation}
\int_{0}^{1}\hat{C}(\hat{r},t)d\hat{r}\leq1\label{eq:CSum}
\end{equation}

The comparison of the dynamics of ionic concentration, versus the
dendrite growth rate indicates that the electrodeposition occurs in
a significantly faster kinetics than dendrites growth:

\[
\frac{\partial\hat{C}}{\partial t}\gg\frac{\partial\hat{\lambda}}{\partial t}
\]

Since the dendrites are the boundary condition for the concentration
development per see, such a distinction implies that the concentration
profile would occur in the quasi-steady state regime in the double
layer region. This is particularly true during stage of instigation
of microstructures, where the nucleation rate is negligible. Therefore
the concentration profile would be obtained by solving the RHS of
Equation \ref{eq:Laplacian} as:

\begin{equation}
\frac{\partial^{2}\hat{C}}{\partial\hat{r}^{2}}+\frac{\kappa}{r_{d}+\kappa\hat{r}}\frac{\partial\hat{C}}{\partial\hat{r}}\approx0\label{eq:SSEq}
\end{equation}

Setting the boundary condition from Equation \ref{eq:BCs} as ${\displaystyle \frac{\partial\hat{C}}{\partial\hat{r}}(0,t)=-\frac{\kappa j}{C_{\infty}D^{+}}}$,
one gets:

\[
\frac{d\hat{C}}{d\hat{r}}\approx\frac{-\kappa r_{d}j}{C_{\infty}D^{+}(r_{d}+\kappa\hat{r})}
\]

Integrating again and having ${\displaystyle \hat{C}(1,t)=1}$ leads
to:

\begin{equation}
\hat{C}(\hat{r},t)\approx1-\frac{r_{d}j}{C_{\infty}D^{+}}\ln\left(\frac{r_{d}+\kappa}{r_{d}+\kappa\hat{r}}\right)\label{eq:Linear}
\end{equation}

For linearization the Equation \ref{eq:Linear} can be re-arranged
as:

\[
\hat{C}(\hat{r},t)\approx1+\frac{r_{d}j}{C_{\infty}D^{+}}\ln\left(1-\frac{\kappa-\kappa\hat{r}}{r_{d}+\kappa}\right)
\]

For the mesoscale dendrite the thickness of the double layer $\kappa$
is negligible relative to the radius of the dendrite $r_{d}$. (i.e.
$\kappa\ll r_{d}$). therefore ${\displaystyle \frac{\kappa-\kappa\hat{r}}{r_{d}+\kappa}\to0}$
and log term can be approximated with the first term of Taylor expansion
as: \footnote{By Taylor expansion $\ln(1+\epsilon)\approx\epsilon-{\displaystyle \cancelto{0}{O(\epsilon^{2})}}$,
where $0<\epsilon\ll1$.}
\begin{align}
\hat{C}(\hat{r},t) & \approx1-\frac{r_{d}j}{C_{\infty}D^{+}}\frac{\kappa(1-\hat{r})}{r_{d}+\kappa}\label{eq:CLinearized}\\
 & \approx1-\frac{\kappa j}{C_{\infty}D^{+}}(1-\hat{r})\nonumber 
\end{align}
Such a linear concentration profile has been addressed in the past
for the flat electrodes as well. \cite{MONROE_03} This profile has
been illustrated in Figure \ref{fig:CProfile} as well. It is obvious
that at the reaction sites ($\hat{r}\rightarrow0$) the concentration
correlates inversely with the ionic flux $j$. In order to have complete
depletion in the reduction sites, we should have the following:

\[
j_{*}=D^{+}\frac{C_{\infty}}{\kappa}
\]

which resembles the flux from the linear concentration distribution
throughout the entire span of double layer and has been expressed
as the critical current density, where the electrode concentration
goes to zero. \cite{Brissot_99_1,Chen_19}

\subsection{Relaxation time}

The Equation \ref{eq:DFreeTFree} in fact resembles the Van Neumann
stability criterion for the typical diffusion equation as: \cite{Isaacson_12}

\begin{equation}
\frac{D^{+}\delta t}{\delta r^{2}}\leq\frac{1}{2}\label{eq:TypicalD}
\end{equation}

the implication is shown in Equation \ref{eq:TimeMeasure} suggests
that the relaxation time correlates with the geometric mean of the
thickness of the double layer $\kappa$ and the outer radius $r_{O}=\kappa+r_{d}$.
The relaxation profile during this time also has been shown in the
Figure \ref{fig:CProfile} where the marginal deviation from the absolutely
uniform concentration distribution (i.e. where $t_{OFF}\to\infty$)
could be due to the round off error as well as truncation error during
the discrete computation.\cite{Milne_53} The geometric mean correlation
for the relaxation time has been addressed before as the \emph{RC
time} of the system for blocking flat electrodes \cite{Bazant_04}
which implies that the regime of relaxation time would vary across
the morphology of the electrodeposits with varying radius of curvature
from atomic scale in the dendritic tips, to the completely flat surface
in smooth areas ($r_{d}\in[r_{atom},\infty]$). Therefore the homogenized
relaxation time would have the following span: \footnote{The counter electrode does not geometry-wise interfere with the double
layer, i.e. $\kappa\ll l$.}

\begin{equation}
\frac{\kappa(\kappa+r_{d})}{D^{+}}\leq t_{OFF}^{opt}\leq\frac{\kappa l}{D^{+}}\label{eq:TimesCompared}
\end{equation}

Nevertheless, the overall relaxation time of the heterogeneous morphology
is determined by the longest relaxation time as the most conservative
case, belonging to the flat zones.

\subsection{Geometry}

As shown in Figure \ref{fig:CurvedDendrite}, the convex boundary
of dendritic interface is exposed to expanded space in the double
layer medium ($r_{O}>r_{d}$). Such geometry alters the dynamics of
concentration gradient relative to flat surface during the pulse-rest
cycle. During the pulse period (i.e. formation of concentration gradient)
the dendritic sites have limited space for the higher feed of ions
from the larges space. Hence, the depletion of concentration occurs
in slower rate, whereas during the relaxation, there will be larger
free domain to diffuse into, relative to flat surface. Therefore the
relaxation occurs with faster rate for convex surfaces. This is also
obvious from Equation \ref{eq:Laplacian} where the term ${\displaystyle \frac{1}{r}\frac{\partial}{\partial r}}$
would alter the concentration dynamics as illustrated in the Table
\ref{tab:CurvatureRole}. The sign of second derivate ${\displaystyle \frac{\partial^{2}C}{\partial r^{2}}}$
is easily discernible from curvature of the concentration profile
in Figure \ref{fig:CProfile}, which is the identical for all morphologies
concerned. \cite{Leger_98,Leger_97}. For convex dendrites, the curvature
term slows down the formation of concentration gradient, whereas it
accelerates the relaxation rate. Following the same phenomenology,
the relaxation in concave surfaces (i.e. pores) occurs at faster dynamics
, as the concentrated atoms have relatively less space to diffuse
into. Respectively the curvature would resist the relaxation for the
pores due to lack of space. Such dynamics translates into the number
of iterations for convergence in our computations.

\begin{table}
\begin{centering}
\begin{tabular}{|c|c|cc||c|cc||c|}
\cline{2-8} \cline{3-8} \cline{4-8} \cline{5-8} \cline{6-8} \cline{7-8} \cline{8-8} 
 & Curvature & \multicolumn{3}{c|}{\textbf{Convex (dendrites)}} & \multicolumn{3}{c|}{\textbf{Concave (pores)}}\tabularnewline
\hline 
Period & ${\displaystyle \frac{\partial^{2}C}{\partial r^{2}}}$ & ${\displaystyle \frac{1}{r}\frac{\partial C}{\partial r}}$ & \multicolumn{2}{c|}{${\displaystyle \frac{\partial C}{\partial t}}$} & ${\displaystyle \frac{1}{r}\frac{\partial C}{\partial r}}$ & \multicolumn{2}{c|}{${\displaystyle \frac{\partial C}{\partial t}}$}\tabularnewline
\cline{1-1} 
Pulse (formation) & $-$ & $+$ & \multicolumn{2}{c|}{Slower} & $-$ & \multicolumn{2}{c|}{Faster}\tabularnewline
\cline{1-1} 
Rest (relaxation) & $+$ & $+$ & \multicolumn{2}{c|}{Faster} & $-$ & \multicolumn{2}{c|}{Slower}\tabularnewline
\hline 
\end{tabular}
\par\end{centering}
\caption{The role of curvature on concentration dynamics versus flat surface.\label{tab:CurvatureRole}}
\end{table}

\begin{figure}
\noindent\begin{minipage}[t]{1\columnwidth}%
\begin{minipage}[c]{0.49\textwidth}%
\begin{center}
\includegraphics[height=0.24\textheight]{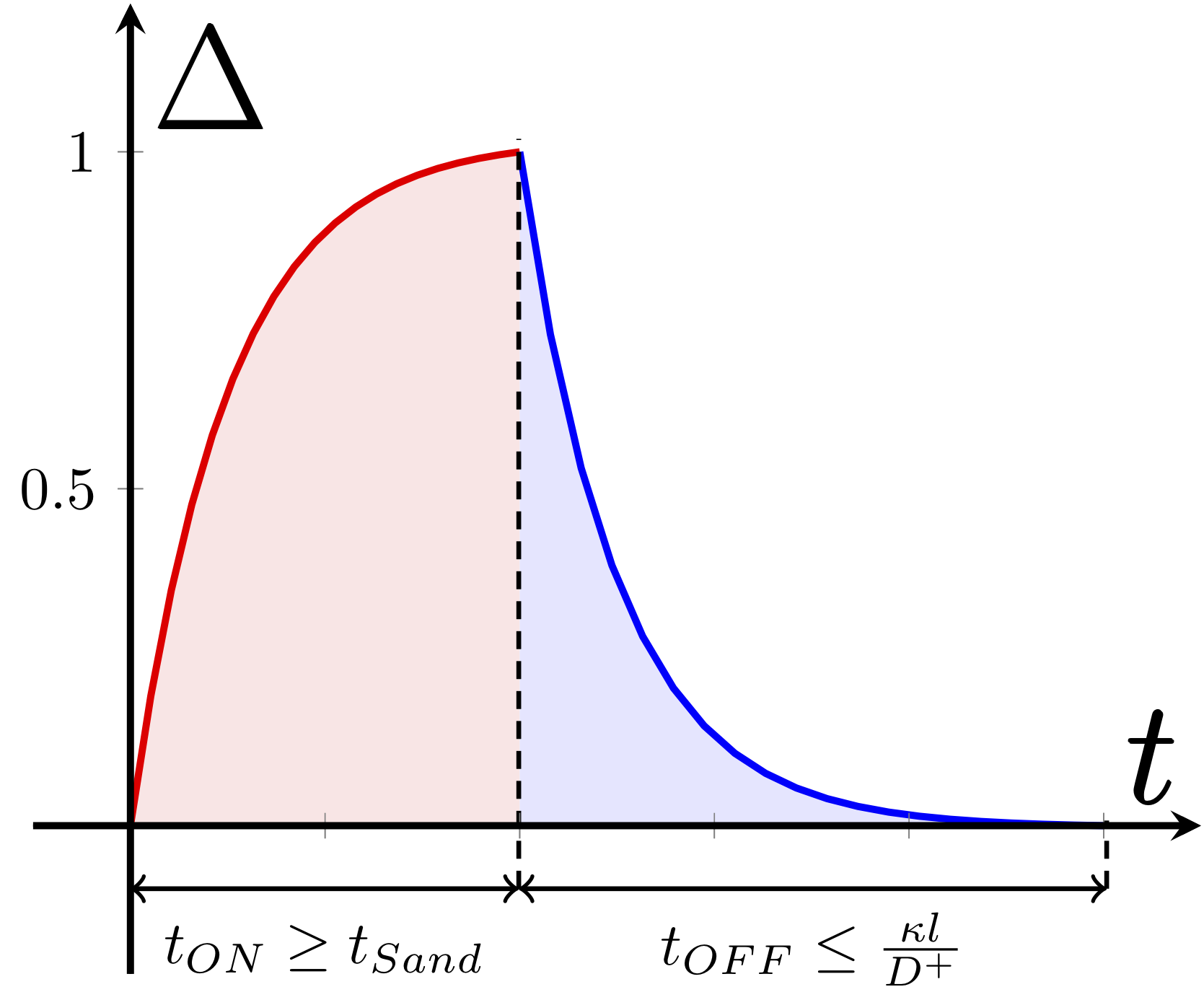}
\par\end{center}
\caption{The time regime for concentration gradient in the convex dendrites.
\label{fig:Delta}}
\end{minipage}\hfill{}%
\begin{minipage}[c]{0.49\textwidth}%
\begin{center}
\includegraphics[height=0.24\textheight]{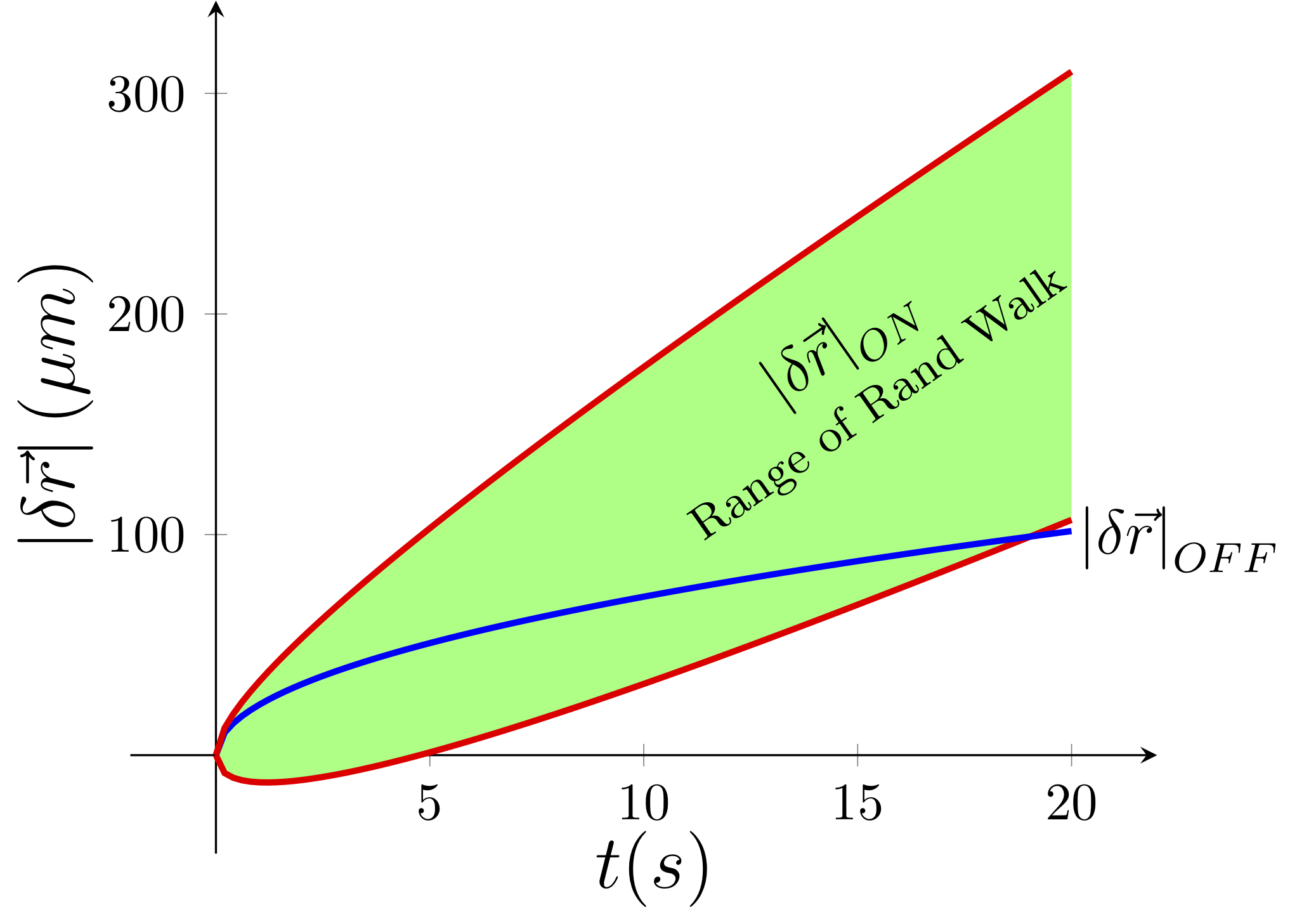}\caption{Extended range of random walk during pulse interval, compared with
the progress of diffusion wave in the rest period.\label{fig:OnOffRange}}
\par\end{center}%
\end{minipage}%
\end{minipage}
\end{figure}

The concentration gradient $\nabla C$ plays the major role for nonuniform
localization of dendritic structures and has a nonlinear behavior
in time. During pulse period, as $\nabla C$ decreases, the rate of
relaxation decreases as well and vanishes when converging to equilibrium
(i.e. uniform profile). We define the depletion measure $\Delta(t)$
for tracking its dynamics as:

\begin{equation}
\Delta(t):=1-\hat{C}_{i}(t)\label{eq:Error}
\end{equation}

where $\hat{C}_{i}(t)$ is the concentration of the interface (i.e.
surface of the dendrite). The variation of depletion measure $\Delta(t)$
during the full pulse-rest period is shown in Figure \ref{fig:Delta},
assuming that the depletion current density meets the critical value
(i.e. $j\geq j_{*}$).\cite{Brissot_99_1} As is discussed in Table
\ref{tab:CurvatureRole}, the formation of such gradient occurs in
a longer time period than the Sand's time \cite{Rosso_07}, whereas
the relaxation occurs faster rate (i.e. shorter time) than the flat
electrode. This is also obvious from Equation \ref{eq:TimesCompared}.\footnote{Obviously based on the geometry of the electrochemical cell. The inter-electrode
distance is the largest dimension amongst all the parameters considered.
Therefore: $r_{d}+\kappa\leq l$.}

Looking at the extended range of diffusion-migration dynamics at provided
more insight to the range of duty cycle. The diffusion length scales
with square root of time ($\delta\vec{r}_{D}\propto t^{\nicefrac{1}{2}}$)
whereas the migration lead scales linearly with it ($\delta\vec{r}_{M}\propto t$).
Therefore, one expects that given a sufficient time $\delta t_{eq}$
, the migration front would take over the diffusion lead. The hypothetical
comparison of the progress of sole-diffusive and sole-migrative waves
is possible from Equations \ref{eq:DiffDis} and \ref{eq:MigDis}
combined with the Einstein relationship ($D^{+}=\mu^{+}RT$ )\cite{Bard_80}:

\begin{equation}
\delta t_{eq}=\frac{2RT}{\mu^{+}|\vec{\textbf{E}}|^{2}}\label{eq:teq}
\end{equation}

where $R$ is gas constant\footnote{R=8.314 $\nicefrac{J}{mol.K}$}
and $T$ is temperature. Closer look at the dynamics of progress during
pulse and rest periods from Equation \ref{eq:OnIneq} leads to the
Figure \ref{fig:OnOffRange}. It is clear that during the initial
moments, the progress in the rest period could be more competitive
with the pulse time. From the Equation \ref{eq:gammaProof}, initially,
the idle ratio $\gamma$ decays exponentially versus the dimension-less
charge period $t_{ON}$. The exponential decay behavior indicates
that relatively shorter amount of idle ratio is needed so that the
diffusion lead would catch up the progress during applied pulse period.
This is also obvious from Equation \ref{eq:gamma}, as the term $\sqrt{\gamma}$
is comparable and in the order of $\gamma$. As the pulse period $t_{ON}$
increases, by neglecting the lower order terms, we reach to the limit
$\gamma\propto t_{ON}$, which directly means $t_{OFF}\propto t_{ON}^{2}$
, therefore for higher applied pulse period $t_{ON}$, the equivalent
idle period $t_{OFF}$ for concentration relaxation has to be significantly
higher. As well in the Equation \ref{eq:Duty} if $t_{ON}$ increase
indefinitely, the correlation of the needed rest period $t_{OFF}$
for the given pulse period $t_{ON}$ will move toward linear relationship
from exponential decay behavior. On the other extreme, the application
of indefinitely high pulse frequency $f$ (i.e. $t_{ON}\rightarrow0$)
might not let the ions reach the reaction sites. Therefore, the \emph{fine-enough}
pulse period would make the applied rest period for the charge relaxation
easier to be competitive with it, as depicted in Figure \ref{fig:SafeUnsafe}.

\section*{List of Symbols}

\begin{minipage}[t]{0.45\columnwidth}%
$f:$ pulse frequency ($Hz$)

$P$ : total period ($s$)

$t_{ON}:$ pulse period ($s$)

$t_{OFF}$: rest period ($s$)

$\delta t$: finite time increment ($s$)

$\hat{\textbf{g}}$ : normal vector in random direction ($[]$)

$\gamma$ : idle ratio ($[]$)

$\mathbf{D}$ : duty cycle ($[]$)

$C$: ionic concentration ($M$)

$\phi$ : electric field ($V$)

$\vec{\textbf{r}}_{D}$: diffusion vector ($m$)

$\vec{\textbf{r}}_{M}$: migration vector ($m$)

$t$ : time ($s$)

$D^{+}$: cationic diffusion coefficient ($m^{2}/s$)

$\vec{E}$ : electric field ($V/m$)

$R$ : gas constant (8.314$j/mol.K$)

$C_{i}$: concentration of the interface ($M$)%
\end{minipage}\hfill{}%
\begin{minipage}[t]{0.45\columnwidth}%
$\mu^{+}$: cationic ionic mobility ($m/V.s$)

$i$ : current density ($mA/cm^{2}$)

$j$ : current flux ($mol/(m^{2}s)$)

$D$ : diameter of the cell ($m$)

$F$ : Faraday's constant ($96.5\text{ }\nicefrac{kC}{mol}$)

$z$: valence number ($[]$)

$\hat{\lambda}$: normalized dendrite measure ($[]$)

$\bar{\lambda}$ : Average normalized dendrite measure ($[]$)

$l$ : inter-electrode distance

$T$ : temperature ($K$)

$r$: radial distance ($m$)

$r_{d}$: radius of curvature of dendrite ($m$)

$r_{O}$: radius of curvature of the outer region ($m$)

$C_{\infty}$ : ambient concentration (i.e. electron-neutral) ($M$)

$\kappa$ : curvature of the interface ($m^{-1}$)%
\end{minipage}

\section{Conclusions}

In this paper, we have performed analytical developments from stochastic
ionic dynamics for the effective suppression of growing dendritic
microstructures during electrodeposition. We defined such square pulse
charging parameters in terms of the range of pulse duty cycle $\mathbf{D}$
and the respective idle time period $t_{OFF}$. Our model considers
the localizations of both ionic concentration and electric field within
the interface of the electrochemical cell, where the nonlinear role
of the dendrite curvature on the relaxation is demonstrated in terms
of cell geometry and the transport property of the electrolyte solution.
The results are useful for estimating the effective charging for dendrite-prone
electrochemical environments, particularly those of involving metallic
electrodes (i.e. lithium, etc.).

\section*{Acknowledgement}

The authors would like to gratefully thank the financial support from
Bill and Melinda Gates Foundation, grant number OPP1192374 and the
BAP support from Bahçe\c{s}ehir University. Additionally, the insightful
discussions during various instances with Dr. Agustin Colussi (Caltech),
Dr. Jaime Marian (UCLA) and Dr. Martin Bazant (MIT) is acknowledged.

\bibliographystyle{unsrt}
\bibliography{/Users/aryanfar/Dropbox/PAPERS/Refs}

\end{document}